%% file: main.tex
\documentclass[sigconf, nonacm]{acmart}

\def\code#1{\texttt{#1}}
\usepackage{makecell}

\newcommand\vldbdoi{XX.XX/XXX.XX}

\newcommand\vldbavailabilityurl{https://github.com/aws-samples/amazon-neptune-samples/tree/master/1g-playground}
\newcommand\vldbpagestyle{plain} 

\begin{document}

\title{Bridging graph data models: RDF,
RDF-star, and property graphs as
directed acyclic graphs [extended abstract]}

\author{Ewout Gelling}
\affiliation{%
  \institution{Eindhoven University of Technology}
}
\email{ewoutgelling1999@gmail.com}
\author{George Fletcher}
\affiliation{%
  \institution{Eindhoven University of Technology}
}
\email{g.h.l.fletcher@tue.nl}
\author{Michael Schmidt}
\affiliation{%
  \institution{Amazon Web Services}
}
\email{schmdtm@amazon.com}

\begin{abstract}

\input{text/abstract.tex}

\end{abstract}

\maketitle

%
%
\pagestyle{\vldbpagestyle}

\ifdefempty{\vldbavailabilityurl}{https://github.com/aws-samples/amazon-neptune-samples/tree/master/1g-playground}{
\vspace{.3cm}
}
%
%

\section{Introduction}

\input{text/chapters/chapter_0_introduction.tex}

\section{Contributions}

\input{text/chapters/chapter_1_contributions.tex}

\section{Background and problem description}

\input{text/chapters/chapter_2_background.tex}

\section{Statement graphs}

\input{text/chapters/chapter_3_statement_graphs.tex}

\section{Concrete Statement graphs}

\input{text/chapters/chapter_4_concrete_statement_graphs.tex}

\section{Interoperability} \label{chapter:interop}
\input{text/chapters/chapter_5_interoperability.tex}

\section{1G Playground}

\input{text/chapters/chapter_6_playground.tex}

\section{Related work}

\input{text/chapters/chapter_7_related_work.tex}

\section{Conclusion and Future Work}

\input{text/chapters/chapter_8_conclusion.tex}

%


\bibliographystyle{ACM-Reference-Format}
\bibliography{sample}

\clearpage

\section{Appendix}

\input{text/appendix}

\end{document}

%% file: text/abstract.tex
Graph database users today face a choice between two technology stacks: the Resource Description Framework (RDF), on one side, is a data model with built-in semantics that was originally developed by the W3C to exchange interconnected data on the Web; on the other side, Labeled Property Graphs (LPGs) are geared towards efficient graph processing and have strong roots in developer and engineering communities. The two models look at graphs from different abstraction layers (triples in RDF vs. edges connecting vertices with inlined properties in LPGs), expose — at least at the surface — distinct features, come with different query languages, and are embedded into their own software ecosystems.

In this short paper, we introduce a novel unifying graph data model called {\it Statement Graphs}, which combines the traits of both RDF and LPG and achieves interoperability at different levels: it (a) provides the ability to manage RDF and LPG data as a single, interconnected graph, (b) supports querying over the integrated graph using any RDF or LPG query language, while (c) clearing the way for graph stack independent data exchange mechanisms and formats. We formalize our new model as directed acyclic graphs and sketch a system of bidirectional mappings between RDF, LPGs, and Statement Graphs. Our mappings implicitly define read query semantics for RDF and LPGs query languages over the unified data model, thus providing graph users with the flexibility to use the query language of their choice for their graph use cases.

As a proof of concept for our ideas, we also present the 1G Playground; an in-memory DBMS built on the concepts of Statement Graphs, which facilitates storage of both RDF and LPG data, and allows for cross-model querying using both SPARQL and Gremlin.




%% file: text/chapters/chapter_0_introduction.tex
The Resource Description Framework (RDF~\cite{rdfw3c}) and Labeled Property Graphs (LPGs) are the two predominant graph data models encountered in today's industrial graph database landscape~\cite{tian2023world}. 

RDF has been developed and standardized as part of the W3C's Semantic Web~\cite{berners2001semantic} initiative. At its core, an RDF dataset is defined as a set of ({\it subject}, {\it predicate}, {\it object}) triples, where each triple represents a fact such as ({\it Alice}, {\it knows}, {\it Bob}). RDF triples can be grouped into containers called {\it named graphs}, and an RDF dataset is defined as a collection of such graphs. On top of RDF, the W3C has developed and standardized higher-level languages that support modeling and inference (RDFS~\cite{rdfsw3c}, OWL~\cite{owlw3c}), as well as a declarative query language called SPARQL~\cite{sparqlw3c}. While not explicitly defined using graph concepts (such as vertices and edges), there is a close connection between RDF and graphs: every triple can be understood as an edge from a {\it subject} node under a given {\it predicate} label to an {\it object} node. In fact, the W3C itself labels RDF a ``graph-based data model''~\cite{rdfw3c} and leverages graphs as an intuitive way to visualize RDF (see e.g.~\cite{rdfprimerw3c}) -- 
yet does not provide a formal mapping between RDF and common graph concepts such as nodes and edges.

In contrast, Labeled Property Graphs are formalized using graph terminology and concepts \cite{AnglesBDFHHLLLM21,pgschema,bonifati2018querying}. The LPG data model consists of sets of vertices and connecting edges, where both vertices and edges can be described through key-value pairs, so-called properties (e.g.~\cite{bonifati2018querying}). In contrast to SPARQL, 
LPG query languages such as Gremlin~\cite{gremlin} and openCypher~\cite{opencypher} have dedicated constructs to access and operate over vertices and edges (edges are called ``relationships'' in openCypher), treating them as first-order concepts. 

One specific feature that makes LPGs a prominent choice for real-world use cases is its built-in support for edge properties; for instance, the distance associated with a {\it route} edge connecting Frankfurt airport (FRA) with London Heathrow (LHR) can be attached to the edge route in form of a property {\it distance -> ``655''}. RDF, where everything is a triple, has no built-in mechanism to attach properties to the ``edge'' (strictly speaking, there is not even a notion of an edge in RDF). The designated mechanism for modeling such scenarios in RDF is the reification vocabulary~\cite{rdfsw3c}, which allows ``statements about statements''. In the example above, it can be used to express that a triple describing a route between two airports has a certain distance, by means of four additional triples:

{\small
\begin{verbatim}
    (FRA, route, LHR), // original triple
    (s, rdf:subject, FRA), (s, rdf:predicate, route),
    (s, rdf:object, LHR), (s, distance, 655)
\end{verbatim}
}

In this example, {\tt s} is a new identifier that represents the underlying triple {\tt (FRA, route, LHR)}; it is defined via pointers to the three position of the triple, using reserved predicates from the RDF namespace, {\tt rdf:subject}, {\tt rdf:predicate}, and {\tt rdf:object}, respectively. The identifier {\tt s} is then used in the last triple to attache the distance. While conceptually sound, RDF reification has been criticized for its verbosity
and poor usability~\cite{hartig2014foundations, nguyen2014don}. Other approaches to ``model around'' lacking edge property support in RDF have similiar limitations: n-ary relations~\cite{naryw3c}, for instance, alter the graph topology and complicate querying; utilizing named graphs for reification (as proposed in ~\cite{trame2013linked}) occupies the graph container, which then can no longer be used for other purposes.  

In response to the (usability) gap for edge properties in RDF, the W3C recently started the RDF-star working group\footnote{\url{https://www.w3.org/groups/wg/rdf-star}}, which aims to establish RDF extensions that provide a concise, user-friendly syntax for expressing edge properties in RDF (and to query them in SPARQL). While the working group seeks to address an important usability aspect of RDF, its outcome will (at best) close an existing gap in one of the standards -- but not overcome the fundamental problem that RDF and LPG are two separate technology stacks that look at graphs from different layers of abstraction. Edge properties are only {\it one amongst many} examples where the stacks differ; they both have very unique strengths (and weaknesses). To give just a few examples, (i) RDF offers great support for global data exchange, publishing, and graph merging (e.g.,~\cite{bizer2008linked}; (ii) LPG query languages, Gremlin and openCypher, offer first-level support for paths as data types, whereas SPARQL comes with built-in support for federation across different endpoints~\cite{sparqlfederationw3c}; (iii) both RDF and LPGs come with their own software ecosystems, developer communities (e.g., Apache Tinkerpop~\cite{gremlin}), and tooling around graph processing.

We argue that, in the end, graph users just want to solve their graph use cases. Having to choose between either of the two models -- and, even worse, being locked into of the two stacks -- stands in direct conflict with their desire to maintain flexibility in a world of changing requirements and emerging opportunities. In order to give these users the flexibility to address their graph problems in the most convenient way, in this paper we are exploring an approach to achieve graph data model interoperability through a novel, overarching data model, that has enough expressive power to capture the RDF, RDF-star, and property graphs data models alike. We next describe two concrete real-world use cases that illustrate the value of such an overarching graph data model.  \\

\noindent \textbf{Use Case: Foodie Travel.} 
A new startup in the airline tickets industry wants to offer their customers the most cost effective way to travel to their destination. They have decided to use a property graph database system as their backing data store. Herein, airports are modeled as nodes, and flights between airports as edges. Each edge carries information about the flight's cost, in the form of an edge property. Use of openCypher path queries allows them to find the most cost-effective route to a customers destination, which may not always be a direct flight. Facing heavy competition, the startup decides to pivot, and cater to a more niche market; that of food enthusiasts. Their new value proposition is to provide the most economical flights to Michelin star rated restaurants. They know that the DBpedia~\cite{mendes2012dbpedia} knowledge graph contains ample information about accredited restaurants, which they aim to integrate with their system. However, herein lies a problem: this data is only accessible as RDF. The inclusion of a triplestore in the architecture would result in additional query overhead and an increase in system complexity. Having both datasets in the same data system would be more practical and performant. Unfortunately, conversion of the RDF data into an LPG format is also ruled-out as an option, because this will make it more challenging to incorporate additional relevant RDF data in the future (for example, a filter for restaurants in buildings which are considered historic). In essence, the startup is in need of a way to combine both RDF and LPG data into a single, queryable structure, where original data is kept in-tact. \\


\noindent \textbf{Use Case: Event Knowledge Graph Analytics.} An {\em Event Knowledge Graph} (EKG) \cite{ekgs_1, ekgs_2} is an LPG based data model relevant to the field of process mining. It describes the logical flow of {\em events} in a business process. Events are modeled as graph nodes and carry a descriptive {\em activity name} such as "Create Invoice" or "Receive Payment". Transitions between events are modeled as directed edges, to which additional information can be linked in the form of {\em entities}. These are separate nodes in the graph, with unique identifiers. For example, the transition between the events "Receive payment" and "Clear invoice" might be associated with an entity containing information about that particular payment. This association is established by referencing the entity identifier in an edge property of a transition. An EKG can ultimately be used to discover the life cycle of a particular entity in the process, where event nodes along a path of transitions that reference said entity are aggregated with an openCypher path query.

Notice that the design of the EKG is affected by a mismatch between the features of the underlying data model and desired query language. The domain calls for the association of entity nodes with event transition edges, a functionality that is not inherently supported in the property graph model. 
This is solved by referencing the entity in an edge property of the transition. However, this indirect link can cause potential data integrity issues down the line. A more natural approach would be to model this domain in RDF-star, in which the entity can be directly linked with the transition through reification. However, this would complicate access to the model, since the SPARQL-star query language does not support the collection of nodes in arbitrary length path traversals.  Again, in essence EKG is in need of a way to combine key features across the RDF and LPG stacks into a single queryable structure.

%% file: text/chapters/chapter_1_contributions.tex
Towards addressing practical use cases such as these, commonly arising across application domains,  in this short paper we make the following contributions:
\begin{itemize}
    \item A formalization of the Statement Graph data model; an overarching data model for RDF, RDF-star, and property graphs.
    \item An open-source proof of concept implementation for Statement Graphs called the `1G Playground'\footnote{\url{https://github.com/aws-samples/amazon-neptune-samples/tree/master/1g-playground}}; an in-memory DBMS that supports both RDF and property graphs, and allows for cross-model querying in SPARQL and Gremlin.
\end{itemize}
Our contributions provide solid foundations and pave the way for graph stack independent management, querying, and exchange of graph data.

%% file: text/chapters/chapter_2_background.tex
There have been different efforts to define interoperability between the {\em surface} data models (RDF, RDF-star, and LPG) through direct mappings\cite{mappings_1,mappings_2,mappings_3,mappings_4}. These primarily make use of the fact that RDF(-star) triples and property graph edges are both 3-ary relations to convert information. However, because the models do not have fully compatible feature sets, these translation are inherently not lossless without introducing additional semantic meaning. Hence, we take a different approach to graph model interoperability: Instead of defining direct mappings between the data models, we raise the three to a common level of abstraction, which we use as an intermediate layer in our inter-data model mappings. \\

We start off by giving formal definitions for the three surface data models. Herein, we will refer to the terms {\em concrete types} and {\em concrete elements}. A {\em concrete type} acts as a building block of some arbitrary data model. Let $\mathcal{E}_1,\ldots,\mathcal{E}_k$ denote concrete types for some $k > 0$. Let an element $e \in \mathcal{E}_1,\ldots,\mathcal{E}_k$ be called a {\em concrete element}. 

An example of a concrete type is the set of all property graph labels $\mathcal{K}$. In which the label $city \in \mathcal{K}$ is an example of a concrete element. We use the following 6 concrete types in our definitions: literals $\mathcal{L}$, IRI's $\mathcal{I}$, blank nodes $\mathcal{B}$, labels $\mathcal{K}$, property names $\mathcal{P}$, and property values $\mathcal{V}$. \\

An \textit{RDF triple} is a tuple $t = (s, p, o)$ where $s \in \mathcal{I} \cup \mathcal{B}$ is called the \textit{subject}, $p \in \mathcal{I}$ is called the \textit{predicate}, and $o \in \mathcal{I} \cup \mathcal{B} \cup \mathcal{L}$ is called the \textit{object}. An \textit{RDF graph} is a finite set consisting of zero or more RDF triples. An \textit{RDF data set} is a collection of the form $\{G_D,(g_i,G_i),\cdots,(g_n,G_n)\}$, for which $0 \leq i \leq n$, where $G_D$ and $G_i$ are both RDF graphs, and $g_i \in \mathcal{I} \cup \mathcal{B}$. The graph $G_D$ is referred to as the {\em default graph}, and the pairs of form $(g_i,G_i)$ are referred as {\em named graphs}\cite{rdfprimerw3c}. Table \ref{table:example_rdf_restaurant} contains an example of an RDF data set in which a DBpedia\cite{mendes2012dbpedia} fragment of a (fictional) restaurant is modeled.

\begin{table}[h]
  \caption{An RDF data set in Turtle\cite{turtle_syntax} syntax containing data about a restaurant.}
  \label{table:example_rdf_restaurant}
  \begin{tabular}{c}
    \toprule
     \textbf{RDF - (default graph)}\\
    \midrule
    \makecell[lc]{
    \code{@prefix ex: <http://example.org/> .} \\
    \code{@prefix dbr: <http://dbpedia.org/resource/> .} \\
    \code{@prefix dbp: <http://dbpedia.org/property/> .} \\
    \code{@prefix dbo: <http://dbpedia.org/ontology/> .} \\
    \code{} \\
    \code{ex:ChezSG} \\
    \code{\hphantom{ex:ChezSG} a dbr:Restaurant ;} \\
    \code{\hphantom{ex:ChezSG} dbo:cuisine "French, classical"@en ;} \\
    \code{\hphantom{ex:ChezSG} dbp:rating "Michelin guide"@en ;} \\
    \code{\hphantom{ex:ChezSG} dbp:city dbr:Paris .} \\
    \code{dbr:Paris} \\
    \code{\hphantom{dbr:Paris} a dbp:city ;} \\
    \code{\hphantom{dbr:Paris} dbp:name "Paris"@en .} \\    }\\
  \bottomrule
\end{tabular}
\end{table}

An \textit{RDF-star triple} is a tuple $t = (s, p, o)$ where $s \in \mathcal{I} \cup \mathcal{B} \cup \{t_1\}$, $p \in \mathcal{I}$, and $o \in \mathcal{I} \cup \mathcal{B} \cup \mathcal{L} \cup \{t_2\}$, in which $t_1$ and $t_2$ are RDF-star triples (adapted from \cite{hartig2014foundations}). Reference cycles are not allowed in an RDF-star triple, i.e., $t \neq t_1,t_2$ nor can $t$ ever be referenced in $t_1$ or $t_2$ recursively. The definitions of an \textit{RDF-star graph} and \textit{RDF-star data set} are then constructed in similar fashion to their RDF counterparts.

\begin{figure}[h]
  \centering
  \includegraphics[width=\linewidth]{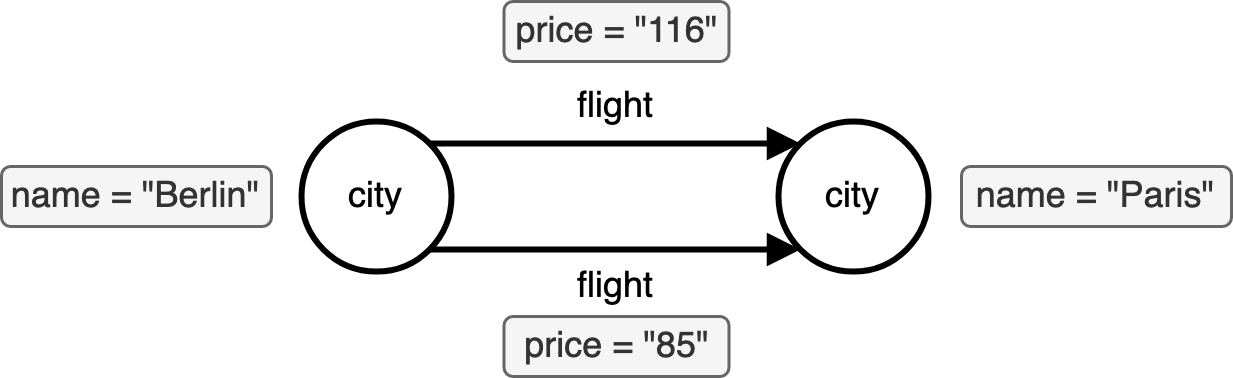}
  \caption{A property graph of flight data.}
  \label{fig:example_flight_data}
\end{figure}

The working definition of a property graph is adapted from \citet{mappings_3}. It is generalized and inclusive, and allows both nodes and edges to be associated with a variable number of labels and properties. A property graph is a tuple: $G = (N, A, P, \delta, \lambda, \sigma, \rho)$ where:

\begin{enumerate}
    \item $N$ is a finite set of nodes, $A$ is a finite set of edges, $P$ is a finite set of properties, and $N, A, P$ are mutually disjoint.
    \item $\delta:A \xrightarrow[]{} (N \times N)$ is a total function that associates each edge in $A$ with a pair of nodes in $N$;
    \item $\lambda:(N \cup A) \to 2^{\mathcal{K}}$ is a total function that associates nodes and edges to a set of labels (possibly empty);
    \item $\sigma:(N \cup A) \to 2^{P}$ is a total function that associates nodes and edges to a set of properties (possibly empty), satisfying that $\sigma(o_1) \cap \sigma(o_2) = \emptyset$ for each pair $o_1,o_2 \in$ $dom(\sigma)$;
    \item $\rho:P \to (\mathcal{P} \times \mathcal{V})$ is a total function that assigns a property name-value pair to each property.
\end{enumerate}

Figure \ref{fig:example_flight_data} shows an example of a property graph in which two flights between Berlin and Paris are modeled as edges. 

%% file: text/chapters/chapter_3_statement_graphs.tex
The Statement Graph data model must be sufficiently flexible to capture the unique traits of all three surface data models simultaneously. This is achieved by making Statement Graphs a semantically agnostic data model, that is to say, by not placing any constraints on data semantics. We formalize statement graphs as directed acyclic graphs (Definition \ref{def_statement_graphs}), in which each internal node has exactly three outgoing edges. These edges have fixed labels, which indicates the ordering of the vertices they point to. Each leaf node of the Statement Graph is associated with a concrete element, and each internal node with a so-called statement identifier (from the concrete type of statement identifiers, $\mathcal{S}$). An internal vertex and its out-neighbours then essentially make up a piece of identifiable information, which we consequently refer to as a~\textit{statement}. Finally, internal nodes can have direct connections to other internal nodes, on the condition that their connecting edge is not labeled $predicate$. Note that we do not imply any sort of physical implementation strategy here, Statement Graphs are a purely logical and conceptual model. Graph users do not interact directly with Statement Graphs, instead, they get the convenience of using the surface data model of their choice.

\begin{definition} \label{def_statement_graphs}

A {\em Statement Graph} $G$ is a directed vertex- and edge-labeled graph $G=(V, E, \tau)$, where:

\begin{itemize} 
    \item $V$ is a finite set of vertices;
    \item $E\subseteq V \times \{subject, predicate, object\} \times V$ is a finite set of labeled edges;
    \item $\tau:V \to \mathcal{S} \cup \mathcal{E}_1 \cup \ldots \cup \mathcal{E}_k$ is a total injective function that labels vertices with statement identifiers or other concrete elements;
    \item every vertex $v \in V$ satisfies the following conditions:
    \begin{enumerate}
        \item If $\tau(v) \in \mathcal{E}_1 \cup \cdots \cup \mathcal{E}_k$ then:
        \begin{enumerate}
            \item $v$ has no outgoing edges,
            \item $v$ has at least 1 incoming edge.
        \end{enumerate}
        \item If $\tau(v) \in \mathcal{S}$ then $v$ has exactly 3 outgoing edges:
        \begin{enumerate}
            \item $(v, subject, v_s)$ and $\tau(v_s) \in \mathcal{S} \cup \mathcal{E}_1 \cup \cdots \cup \mathcal{E}_k$,
            \item $(v, predicate, v_p)$ and $\tau(v_p) \in \mathcal{E}_1 \cup \cdots \cup \mathcal{E}_k$,
            \item $(v, object, v_o)$ and $\tau(v_o) \in \mathcal{S} \cup \mathcal{E}_1 \cup \cdots \cup \mathcal{E}_k$.
        \end{enumerate}
    \end{enumerate}
    \item $G$ is acyclic, i.e., there is no path in $E$ from a vertex to itself. 
 \end{itemize}
\end{definition}

Figures 2 and 3 each show an example of a Statement Graph. In these figures, $subject$ edges are represented as hollow arrows, $predicate$ edges as double arrows, and $object$ edges as single arrows. Note that we could have taken an alternative approach to formalizing Statement Graphs,  e.g., as a class of nested hypergraphs \cite{ubergraphs}.
We adopt the view of directed acyclic graphs, for ease of presentation.




%% file: text/chapters/chapter_4_concrete_statement_graphs.tex
In this section, we introduce concise Statement Graph fragments for our surface data models, which we refer to as Statement Graph {\em Images}. These are designed to be minimal, i.e., each Image captures the traits of its associated surface data model, and nothing more. Lossless, bidirectional mappings exist between the Statement Graph Images and their corresponding surface data models, which shows that they are functionally equivalent. This system of mappings has been omitted for brevity, but is included in the Appendix of the full version of this extended abstract\cite{full_version_github}.

\subsection{RDF data}
In the following definitions we refer to the notion of vertex \textit{isomorphism} within Statement Graphs. We consider two internal vertices to be isomorphic if they have outgoing edges to the same vertices; they are essentially making the "same" statement. This allows RDF-(star) set semantics to be expressed in Statement Graphs. 

The traits of RDF-star are captured in an \textit{RDF-star-Image} (Definition \ref{rdf_star_image}) by distinguishing between two groups of statements: triple statements and membership statements. A triple statement captures the semantics of an RDF-star triple, reification is herein accommodated by allowing statement identifiers to be referenced in $subject$ and $object$ position (see constraints \ref{rdf_star_image_1}, \ref{rdf_star_image_2}, and \ref{rdf_star_image_3}). A membership statement assigns a triple statement to either a graph name, or the default graph identifier, $G_D \in \mathcal{I}$. The constant concrete element $in \in \mathcal{I}$ is used in $predicate$ position of these statements. The final constraint (\ref{rdf_star_image_4}), first lays out the structure of a membership statement (\ref{rdf_star_image_mem_constraint}), then asserts that each triple statement is referenced in the context of at least one other statement (\ref{rdf_star_image_non_mem_constraint}), and ensures statement uniqueness by disallowing vertex {\em isomorphism} (\ref{rdf_star_iso}).

\begin{definition} \label{rdf_star_image}
A Statement Graph is an {\em RDF-star-Image} when:
\begin{enumerate}
    \item For all edges $(v, subject, v_s) \in E$, $\tau(v_s) \in \mathcal{S} \cup \mathcal{I} \cup \mathcal{B}$; \label{rdf_star_image_1}
    \item For all edges $(v, predicate, v_p) \in E$, $\tau(v_p) \in \mathcal{I}$; \label{rdf_star_image_2}
    \item For all edges $(v, object, v_o) \in E$, $\tau(v_o) \in \mathcal{S} \cup \mathcal{I} \cup \mathcal{B} \cup \mathcal{L}$;   \label{rdf_star_image_3}
    \item For every internal vertex $v \in V$, let $v_s, v_p,v_o$ be vertices with incoming $subject$, $predicate$, $object$ edges from $v$, the following constraints hold: \label{rdf_star_image_4}
    \begin{enumerate}
        \item if $\tau(v_p) = in$ then $\tau(v_s) \in \mathcal{S}$ and $\tau(v_o) \in \mathcal{I} \cup \mathcal{B}$ and $v$ has no incoming edges; \label{rdf_star_image_mem_constraint}
        \item if $\tau(v_p) \in \mathcal{I} \setminus in$ then $v$ has at least 1 incoming edge; \label{rdf_star_image_non_mem_constraint}
        \item There is no $v' \in V$ that is {\em isomorphic} to $v$. \label{rdf_star_iso}
    \end{enumerate}
\end{enumerate}
\end{definition}

\begin{figure}[h]
  \centering
  \includegraphics[width=0.95\linewidth]{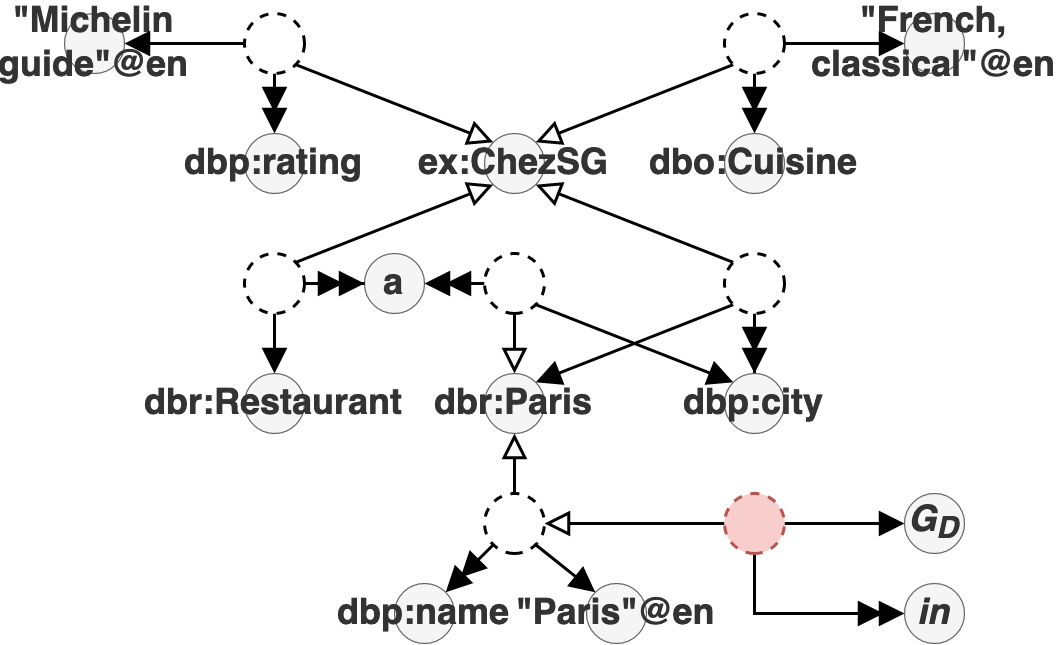}
  \caption{Statement Graph of restaurant data. This is the RDF-Image of the RDF data set given in Table \ref{table:example_rdf_restaurant}.}
  \label{fig:statement_graph_restaurant_data}
\end{figure}

An \textit{RDF-Image} (Definition \ref{rdf_image}) is a special case of RDF-star-Image in which internal nodes can only be referenced in the context of a membership statement. Consequently, each RDF-Image must satisfy two additional constraints on top of those that it inherits, these are: (\ref{rdf_image_reify}) Internal nodes can never reference others in $object$ position, and (\ref{rdf_image_membership}) every internal node must have at least one incoming edge from a membership statement (i.e. be part of at least one graph). Figure \ref{fig:statement_graph_restaurant_data} shows a Statement Graph that is the RDF-Image of the data set from Table \ref{table:example_rdf_restaurant}. In this figure, the node marked in red is a membership statement, all other membership statements are omitted to avoid cluttering.

\begin{definition} \label{rdf_image}
A Statement Graph is an {\em RDF-Image} when it is an RDF-star-Image and:
\begin{enumerate}
    \item For all edges $(v, object, v_o) \in E$, $\tau(v_o) \in \mathcal{I} \cup \mathcal{B} \cup \mathcal{L}$;  \label{rdf_image_reify}
    \item For every internal vertex $v \in V$, let $v_s, v_p,v_o$ be vertices with incoming $subject$, $predicate$, $object$ edges from $v$, the following constraint holds: \label{rdf_image_membership}
    \begin{enumerate}
        \item if $\tau(v_p) \in \mathcal{I} \setminus in$ then $\tau(v_s) \in \mathcal{I} \cup \mathcal{B}$ and $\tau(v_o) \in \mathcal{I} \cup \mathcal{B} \cup \mathcal{L}$ and $v$ has at least 1 incoming edge. 
    \end{enumerate}
\end{enumerate}
\end{definition}

\subsection{LPG data}
\begin{figure}[h]
  \centering
  \includegraphics[width=0.75\linewidth]{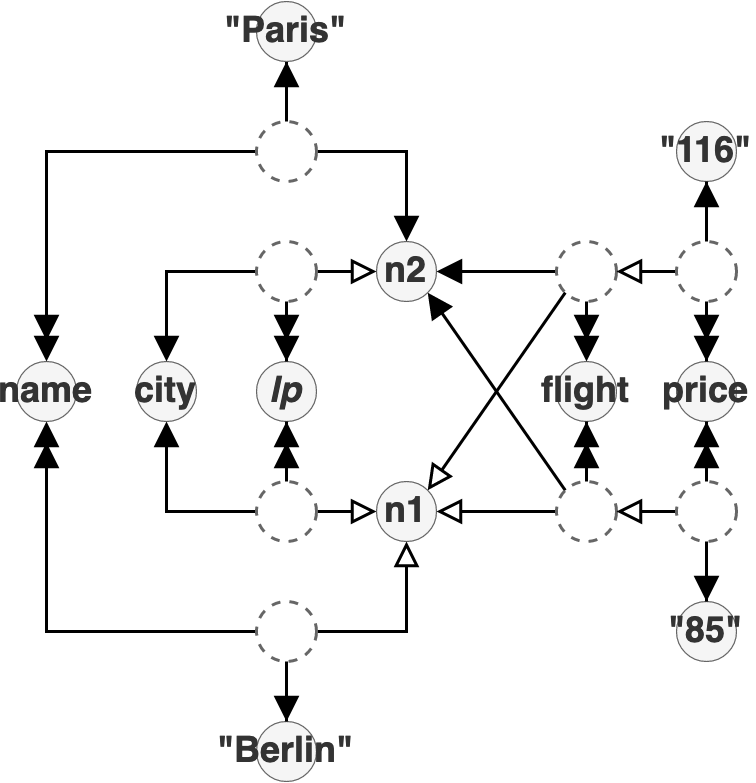}
  \caption{Statement Graph of flight data. This is the LPG-Image of the property graph from Figure \ref{fig:example_flight_data}}
  \label{fig:statement_graph_flight_data}
\end{figure}

Let $\mathcal{N}$ be the concrete type of property graph node identities. For simplicity, we assume that each property graph node is associated with a concrete element from this set. The traits of a property graph are captured in an \textit{LPG-Image} (Definition \ref{pg_image}) by distinguishing between three groups of statements: Edge statements (\ref{pg_image_edge}), node label statements (\ref{pg_image_node_labels}), and property statements (\ref{pg_image_property}). The constant concrete element $l_p \in 2^{\mathcal{K}}$ is used in $predicate$ position of node label statements. Figure \ref{fig:statement_graph_flight_data} shows a Statement Graph that is the LPG-Image of the property graph from Figure \ref{fig:example_flight_data}. In this figure, $n1$ and $n2$ are elements from $\mathcal{N}$.

\begin{definition} \label{pg_image}
A Statement Graph is an {\em LPG-Image} when:
\begin{enumerate}
    \item For all edges $(v, subject, v_s) \in E$, $\tau(v_s) \in \mathcal{N} \cup \mathcal{S}$;
    \item For all edges $(v, predicate, v_p) \in E$, $\tau(v_p) \in \mathcal{P} \cup 2^{\mathcal{K}}$;
    \item For all edges $(v, object, v_o) \in E$, $\tau(v_o) \in \mathcal{V} \cup \mathcal{N} \cup 2^{\mathcal{K}}$;  
    \item For every internal vertex $v \in V$, let $v_s, v_p,v_o$ be vertices with incoming $subject$, $predicate$, $object$ edges from $v$, precisely one of the following constraints hold:
    \begin{enumerate}
        \item if $\tau(v_p) \in 2^{\mathcal{K}} \setminus l_p$ then $\tau(v_s) \in \mathcal{N}$ and $\tau(v_o) \in \mathcal{N}$; \label{pg_image_edge}
        \item if $\tau(v_p) = l_p$ then $\tau(v_s) \in \mathcal{N}$ and $\tau(v_o) \in 2^{\mathcal{K}}$ and $v$ has no incoming edges; \label{pg_image_node_labels}
        \item if $\tau(v_p) \in \mathcal{P}$ then $\tau(v_s) \in \mathcal{S} \cup \mathcal{N}$ and $\tau(v_o) \in \mathcal{V}$ and $v$ has no incoming edges. \label{pg_image_property}
    \end{enumerate}
\end{enumerate}
\end{definition}

\subsection{OneGraph data}

A {\em OneGraph-Image} (Definition \ref{og_image}) is a Statement Graph comprised of a combination of RDF-, RDF-star-, and LPG-Images. It is used in the next section to simplify reasoning about interoperability. The name `OneGraph-Image' is a reference to the OneGraph (1G) vision of \citet{onegraph}. As shown in \cite{ewoutgthesis}, our formalization of Statement Graphs, and more specifically that of OneGraph-Images, acts as a concrete implementation of the OneGraph vision.

\begin{definition} \label{og_image}
A Statement Graph $G$ is a {\em OneGraph-Image} when each component of $G$ is an RDF-star-Image or LPG-Image, disregarding constraints on isomorphism of vertices.
\end{definition}

%% file: text/chapters/chapter_5_interoperability.tex
We have shown that Statement Graphs act as a common abstraction layer for the surface data models, which allows their instances to be aggregated under a single structure. In this section, we focus on how Statement Graphs are able to facilitate their interoperability. We convey this through a series of mappings that show how a OneGraph-Image can be converted into an RDF-, RDF-star-, or LPG-Image. These mappings implicitly define cross-model read query semantics for the surface data models. For example, an RDF Data set can be queried in Gremlin by subsequently converting it into an RDF-Image, LPG-Image, and property graph.

The mappings make use of the following total functions that relate naturally similar concrete types from the RDF and property graph worlds. They are: $literalvalue: \mathcal{L} \mapsto \mathcal{V}$, $blank2node: \mathcal{B} \mapsto \mathcal{N}$, $IRI2node: \mathcal{I} \mapsto \mathcal{N}$, $IRI2prop: \mathcal{I} \mapsto \mathcal{P}$, and $IRI2label: \mathcal{I} \mapsto 2^{\mathcal{K}}$. There is also an inverse to each of these that achieves the opposite goal, e.g., $value2literal: \mathcal{V} \mapsto \mathcal{L}$, etc. These functions are used to convert the concrete elements associated with leaf nodes of the given Statement Graph. Their entries act as parameters to the mappings. For example, an entry in $literal2value$, relating the literal \texttt{"Paris"@en} with property value \texttt{"Paris"}.

In the below translations, vertex removals are recursive, i.e., when a vertex is removed, all vertices with outgoing edges to that vertex are also removed.

\subsection{OneGraph-Image to RDF-(star)-Image}

The following mapping translates the OneGraph-Image $G=(V,E, \tau)$ into an RDF-Image if the optional step (\ref{og_to_rdf_optional}) is taken, or an RDF-star-Image if the optional step is skipped.

\begin{enumerate}
    \item For each leaf node $v \in V$, convert $\tau(v)$ using the appropriate type conversion function if needed, merge nodes with equivalent values for $\tau$;
    \item Add vertices $v_{in}$ and $v_{G_{D}}$ to $V$, with $\tau(v_{in}) = in$ and $\tau(v_{G_{D}}) = G_D$, if such vertices did not already exist in $V$;
    \item For every internal vertex $v \in V$, add vertex $v_{st}$ to $V$, $\tau(v_{st}) \in \mathcal{S}$, add edges  edges $(v_{g}, subject, v),(v_{g},predicate,in)$, and $(v_{g},object,v_{G_{D}})$ to $E$;
    \item For every internal vertex $v \in V$, remove vertices in $V$ that are isomorphic to $v$;
    \item (Optional) For every internal vertex $v \in V$, let $v_s, v_p,v_o$ be vertices with incoming $subject$, $predicate$, $object$ edges from $v$. If either $v_s$ or $v_o$ is an internal vertex and $\tau(v_{p}) \neq in$, then remove $v$; \label{og_to_rdf_optional}
    \item Remove vertices in $V$ without incoming or outgoing edges.
\end{enumerate}

The RDF-Image of the RDF data set of Table 
\ref{table:example_rdf_restaurant}
obtained via this mapping is visualized in Figure \ref{fig:statement_graph_restaurant_data}.

\subsection{OneGraph-Image to LPG-Image}

The following mapping translates the OneGraph-Image $G=(V,E, \tau)$ into an LPG-Image.

\begin{enumerate}
    \item For each leaf node $v \in V$, convert $\tau(v)$ using the appropriate type conversion function if needed, merge nodes with equivalent values for $\tau$. When encountering a $\tau(v) \in \mathcal{I}$:
    \begin{enumerate}
        \item if $v$ has an incoming $subject$ or $object$ edge and no incoming $predicate$ edge; Set $\tau(v)$ = $IRI2node(\tau(v))$
        \item if $v$ has an incoming $predicate$ edge and no incoming $subject$ or $object$ edge; Set $\tau(v)$ = $IRI2label(\tau(v))$
        \item if $v$ has an incoming $subject$ or $object$ edge and an incoming $predicate$ edge;
        \begin{enumerate}
            \item Add node $\overline{v}$ to $V$, let $\tau(\overline{v}) = IRI2node(\tau(v))$;
            \item Let $\tau(v) = IRI2label(\tau(v))$;
            \item Let all $predicate$ edges point to $v$, let all $subject$ and $object$ edges point to $\overline{v}$.
        \end{enumerate}
    \end{enumerate}
    \item For each internal vertex $v \in V$, let $v_s,v_p,v_o$ be vertices with incoming $subject$, $predicate$, $object$ edges from $v$:
    \begin{enumerate}
        \item If $v$ has an incoming edge from some vertex $\overline{v}$, then remove $\overline{v}$ if one of the following conditions hold:
        \begin{enumerate}
            \item The incoming edge to $v$ is not labeled $subject$
            \item $\tau(v_o) \notin \mathcal{N}$
            \item $\overline{v}$ does not have an $object$ edge to a vertex $\overline{v_o}$ with $\tau(\overline{v_o}) \in \mathcal{V}$.
        \end{enumerate} 
    \end{enumerate}
    \item Remove vertices in $V$ without incoming or outgoing edges.
\end{enumerate}

The LPG-Image of the property graph of Figure 
\ref{fig:example_flight_data}
obtained via this mapping is visualized in Figure \ref{fig:statement_graph_flight_data}.

%% file: text/chapters/chapter_6_playground.tex
\begin{figure}[h]
  \centering
  \includegraphics[width=0.7\linewidth]{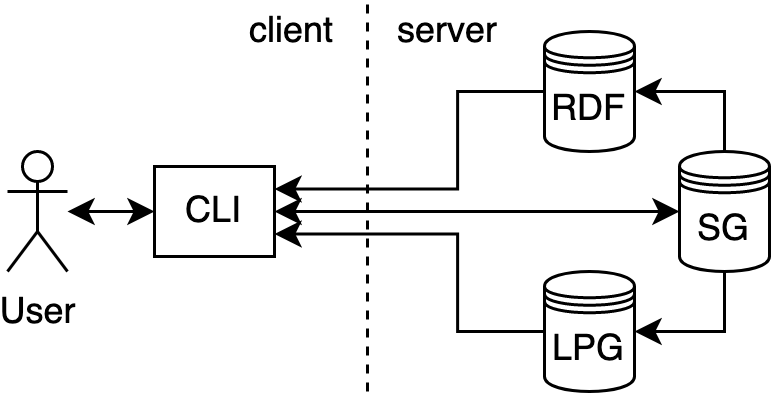}
  \caption{Architecture of the 1G Playground}
  \label{fig:1g_playground_architecture}
\end{figure}

We next describe our open source implementation of Statement Graphs, which we call the ``1G Playground'', highlighting the OneGraph vision which inspired this work. The open source 1G Playground\footnote{\url{https://github.com/aws-samples/amazon-neptune-samples/tree/master/1g-playground}} complements the foundational work in this paper with a practical, demonstrable product that illustrates the key ideas of Statement Graphs and serves as a proof of concept. It consists of two components that adhere to the traditional client-server model. On the client side we have a REPL environment akin to psql\footnote{\url{https://www.postgresql.org/docs/current/app-psql.html}}, while the server side acts as a simple in-memory DBMS (see Figure \ref{fig:1g_playground_architecture}). The server exposes a REST interface with endpoints for the following functionalities: data loading, data retrieving, querying, and modification of current settings. Data can be loaded and retrieved in various different RDF and property graph formats, such as: N-Quads\cite{nquadsw3c}, Turtle\cite{turtle_syntax}, Graphson\footnote{\url{https://tinkerpop.apache.org/docs/3.4.1/dev/io/\#graphson}}, and many more. The Playground also comes with a novel shared serialization format that allows users to export their RDF and LPG data in a unified `1G' syntax (accessible using flag \textit{-og}).

The server component of the playground is made up of a single OneGraph-Image (indicated by SG in the figure), which is accompanied by two derivative data stores; one for LPG (backed by TinkerPop) and one for RDF (backed by RDF4J\footnote{\url{https://rdf4j.org/}}). The user can not modify these data stores directly, only the Statement Graph itself can be written to. RDF and LPG input first gets converted into its corresponding Image before being added to the central graph. The translations described in Section \ref{chapter:interop} are then used to reload both derivative data stores from the updated Statement Graph. The playground allows the user to configure the $IRI2label$, $label2IRI$, $IRI2prop$ and $prop2IRI$ functions, which gives them a certain level of control over these translations. All other concrete type conversions (such as $literal2value$) are performed implicitly. This architecture allows the playground to evaluate SPARQL and Gremlin read queries on the Statement Graph, without having to implement a custom query engine from scratch.

The playground includes a set of interactive demo scenarios which we have developed to help users understand the tool, and to illustrate Statement Graphs and their capabilities. There are three scenarios that can be explored using the \textit{scenario} command, more info about this can be found on the GitHub page.

%% file: text/chapters/chapter_7_related_work.tex
\noindent {\bf Mapping between RDF, RDF-star, and labeled property graphs.}
In recent years there has been quite some effort on designing direct 
model-to-model mappings between RDF, RDF-star, and LPGs,  e.g., \cite{mappings_1,mappings_2,mappings_3,mappings_4}.  Our work differs from and contributes to this body of work by providing an overarching data model supporting uniform translations between all of these surface data models.  Furthermore, taking the data model perspective provides a more fundamental understanding of what graphs are and what is common and different between graphs modeled as RDF, RDF-star, and LPG (and beyond).

\smallskip

\noindent {\bf Unifying (graph) data models.}
The study of graph modeling (e.g., \cite{bonifati2018querying,survey,hypgergraphdb}) and unifying ``meta'' data models  (e.g., \cite{AtzeniT93,AtzeniGC09}) are classic themes in data management research.   Most closely related to our investigation are the OneGraph \cite{onegraph} and Multilayer Graph \cite{multilayer_graphs} proposals, which also take a statement-centric approach to unifying RDF, RDF-star, and LPG.  Our approach complements this work by taking a novel view of statement graphs as directed acyclic graphs, which are familiar, intuitive, and easy to manipulate and reason about.




%% file: text/chapters/chapter_8_conclusion.tex
We have argued that graph users just want to solve their graph use cases, and having to bridge multiple graph data models and graph stacks, as is often the case in contemporary graph work, stands in direct conflict with this desire.  In this paper we have developed solid foundations on which to design and engineer graph systems which resolve this conflict by bridging the worlds of RDF and property graphs, with Statement Graphs, an intuitive and simple unifying graph data model.  We have also described our open source toolkit, 1G Playground, which demonstrates the feasibility and usefulness of Statement Graphs.

Many interesting research challenges arise from our work.  A main focus here can be development and study of the broader framework around Statement Graph-based systems, e.g.,  query languages and query rewriting,  schema and constraint languages and enforcement methods, indexing strategies, and serialization formats.   A concrete way forward on many of these lines of future work could be through further development of the open source 1G Playground.

%% file: text/appendix.tex
\subsection{Property graph to Statement Graph}

The following mapping can be applied to property graph $PG=(N,A,P,\delta,\lambda,\sigma,\rho)$ to obtain an LPG-Image $G=(V,E,\tau)$

\begin{enumerate}
    \item For every $n \in N$:
    \begin{enumerate}
        \item Add vertices $v_s$, $v_p$, $v_o$, and $v_{st}$ to $V$, let $\tau(v_s) = n$, $\tau(v_p) = l_p$, $\tau(v_o) = \lambda(n)$ and let $\tau(v_{st})$ be an unused element from $\mathcal{S}$
        \item Add edges $(v_{st}, subject, v_{s})$, $(v_{st}, predicate, v_p)$, \\ and $(v_{st}, object, v_o)$ to $E$. 
        \item For every property $(p_{name}, p_{value}) \in \rho(\sigma(n))$:
        \begin{enumerate}
            \item Add vertices $\overline{v_p}$, $\overline{v_o}$ and $\overline{v_{st}}$ to $V$, let $\tau(\overline{v_p}) = l_p$, $\tau(\overline{v_o}) = \lambda(n)$ and let $\tau(\overline{v_{st}})$ be an unused element from $\mathcal{S}$
            \item Add edges $(\overline{v_{st}}, subject, v_s)$, $(\overline{v_{st}}, predicate, \overline{v_p})$, and \\ $(\overline{v_{st}}, object, \overline{v_o})$ to $E$.    
        \end{enumerate}
    \end{enumerate}
    \item For every $e \in A$, where $\delta(e) = (n_{1}, n_{2})$:
    \begin{enumerate}
        \item Add vertices $v_s$, $v_p$, $v_o$, and $v_{st}$ to $V$, let $\tau(v_s) = n_1$, $\tau(v_p) = \lambda(e)$, $\tau(v_o) = n_2$ and let $\tau(v_{st})$ be an unused element from $\mathcal{S}$
        \item Add edges $(v_{st}, subject, v_{s})$, $(v_{st}, predicate, v_p)$, \\ and $(v_{st}, object, v_o)$ to $E$. 
        \item For every property $(p_{name}, p_{value}) \in \rho(\sigma(e))$:
        \begin{enumerate}
            \item Add vertices $\overline{v_p}$, $\overline{v_o}$ and $\overline{v_{st}}$ to $V$, let $\tau(\overline{v_p}) = l_p$, $\tau(\overline{v_o}) = \lambda(n)$ and let $\tau(\overline{v_{st}})$ be an unused element from $\mathcal{S}$
            \item Add edges $(\overline{v_{st}}, subject, v_{st})$, $(\overline{v_{st}}, predicate, \overline{v_p})$, and \\ $(\overline{v_{st}}, object, \overline{v_o})$ to $E$.  
        \end{enumerate}
    \end{enumerate}
    \item Merge vertices in $V$ with equal values for $\tau$;
\end{enumerate}

\subsection{Statement Graph to Property graph}

The following mapping can be applied to LPG-Image $G=(V,E,\tau)$ to obtain a property graph $PG=(N,A,P,\delta,\lambda,\sigma,\rho)$.

\begin{enumerate}
    \item Let $V_{st}$ be the set of internal vertices in $V$. When iterating $V_{st}$, let $v_s$, $v_p$, $v_o$ be vertices with incoming $subject$, $predicate$, $object$ edges from a $v \in V_{st}$  
    \item For every $v \in V_{st}$ for which $\tau(v_p) = l_p$:
    \begin{enumerate}
        \item Let node $n = \tau(v_s)$, add $n$ to $N$, set $\lambda(n) = \tau(v_s)$.
    \end{enumerate}
    \item For every $v \in V_{st}$ for which $\tau(v_s) \in \mathcal{N} \wedge \tau(v_o) \in \mathcal{N}$:
    \begin{enumerate}
        \item Let node $n_1 = \tau(v_s)$ and $n_2 = \tau(v_o)$, add edge $e=(n_1,n_2)$ to $A$, set $\lambda(e) = \tau(v_p)$.
    \end{enumerate}
    \item For every $v \in V_{st}$ for which  $\tau(v_o) \in \mathcal{V}$:
    \begin{enumerate}
        \item If $\tau(v_s) \in \mathcal{N}$, let node $n = \tau(v_s)$, add property $p$ to $P$ and entries $\sigma(n) = p$ and $\rho(p) = (\tau(v_p),\tau(v_o))$
        \item If $\tau(v_s) \in \mathcal{S}$, obtain edge $e$ that was previously created for $v_s$, add property $p$ to $P$ and entries $\sigma(e) = p$ and $\rho(p) = (\tau(v_p),\tau(v_o))$
    \end{enumerate}
\end{enumerate}

\subsection{RDF data set to Statement Graph}

The following mapping can be applied to RDF data set \\
$D=\{G_D,(g_i,G_i),\cdots,(g_n,G_n)\}$ to obtain an
RDF-Image $G=(V,E,\tau)$

\begin{enumerate}
    \item Add $v_{in}$ to $V$, let $\tau(v_{in}) = in$
    \item For all elements $g \in D$:
    \begin{enumerate}
        \item Let $(name,graph) = (g_i,G_i)$ if $g$ is a named graph, or let $(name,graph) = (G_{default},G_D)$ if $g$ is the default graph;
        \item Add a new vertex $v_{graph}$ to $G$, let $\tau(v_{graph}) = name$;
        \item For all RDF triples $(s,p,o) \in graph$:
        \begin{enumerate} 
            \item Add $v_s$ to $G$, let $\tau(v_s) = s$;
            \item Add $v_p$ to $G$, let $\tau(v_p) = p$;
            \item Add $v_o$ to $G$, let $\tau(v_o) = o$;
            \item Add $v_{st}$ to $G$, let $\tau(v_{st})$ be an unused element from $\mathcal{S}$
            \item Add edges $(v_{st},subject,v_s)$, $(v_{st},predicate,v_p)$, \\ and $(v_{st},object,v_o)$ to $E$;
            \item Add $v_{mem}$ to $G$, let $\tau(v_{mem})$ be an unused element from $\mathcal{S}$
            \item Add edges $(v_{mem},subject,v_{st})$, $(v_{mem},predicate,v_{in})$ \\ and $(v_{mem},object,v_{graph})$ to $E$;
        \end{enumerate}
    \end{enumerate}
    \item Merge all isomorphic vertices;
    \item Merge vertices in $V$ with equal values for $\tau$.
\end{enumerate}

\subsection{Statement Graph to RDF data set}

The following mapping can be applied to RDF-Image $G=(V,E,\tau)$ to obtain an RDF data set $D=\{G_D,(g_i,G_i),\cdots,(g_n,G_n)\}$, let $D$ contain only the default graph.

\begin{enumerate}
    \item Let $V_{st}$ be the set of internal vertices in $V$. When iterating $V_{st}$, let $v_s$, $v_p$, $v_o$ be vertices with incoming $subject$, $predicate$, $object$ edges from a $v \in V_{st}$;
    \item For every $v \in V_{st}$ for which $\tau(v_p) \neq in$:
    \begin{enumerate}
        \item Create RDF triple $t=(s,p,o)$;
        \item Let $s=\tau(v_s)$, $p=\tau(v_p)$, $o=\tau(v_o)$;
        \item Let $V_{mem}$ be the set of vertices with outgoing edges to $v$;
        \item For all $v_{mem} \in V_{mem}$, let $\overline{v_o}$ be the vertex with incoming $object$ edge from $v_{mem}$:
        \begin{enumerate}
            \item Let $name = \tau(\overline{v_o})$;
            \item If $name = G_D$, add the triple $t$ to the default graph in $D$.
            \item If $name \neq G_D$, add the triple $t$ to the named graph named $name$ in $D$, if no such graph exists create a new one.
        \end{enumerate}
    \end{enumerate}
\end{enumerate}

\subsection{RDF-star data set to Statement Graph}

The following mapping can be applied to RDF-star data set $D=\{G_D,(g_i,G_i),\cdots,(g_n,G_n)\}$ to obtain an RDF-star-Image $G=(V,E,\tau)$

\begin{enumerate}
    \item Add $v_{in}$ to $V$, let $\tau(v_{in}) = in$
    \item For all elements $g \in D$:
    \begin{enumerate}
        \item Let $(name,graph) = (g_i,G_i)$ if $g$ is a named graph, or let $(name,graph) = (G_{default},G_D)$ if $g$ is the default graph;
        \item Add a new vertex $v_{graph}$ to $G$, let $\tau(v_{graph}) = name$;
        \item For all RDF-star triples $(s,p,o) \in graph$:
        \begin{enumerate} 
            \item If $s$ is an RDF-star triple, then execute steps (i), (ii), (iii), (iv), and (v) on $t$. Let $v_s$ be the internal vertex obtained in step (v);
            \item Else, add $v_s$ to $G$, let $\tau(v_s) = s$;
            \item Add $v_p$ to $G$, let $\tau(v_p) = p$;
            \item If $o$ is an RDF-star triple, then execute steps (i), (ii), (iii), (iv), and (v)  on $t$. Let $v_o$ be the internal vertex obtained in step (v);
            \item Else, add $v_o$ to $G$, let $\tau(v_o) = o$;
            \item Add $v_{st}$ to $G$, let $\tau(v_{st})$ be an unused element from $\mathcal{S}$;
            \item Add edges $(v_{st},subject,v_s)$, $(v_{st},predicate,v_p)$, \\ and $(v_{st},object,v_o)$ to $E$;
            \item Add $v_{mem}$ to $G$, let $\tau(v_{mem})$ be an unused element from $\mathcal{S}$;
            \item Add edges $(v_{mem},subject,v_{st})$, $(v_{mem},predicate,v_{in})$, \\ and $(v_{mem},object,v_{graph})$ to $E$;
        \end{enumerate}
    \end{enumerate}
    \item Merge all isomorphic vertices;
    \item Merge vertices in $V$ with equal values for $\tau$.
\end{enumerate}

\subsection{Statement Graph to RDF-star data set}

The following mapping can be applied to RDF-star-Image $G=(V,E,\tau)$ to obtain an RDF-star data set $D=\{G_D,(g_i,G_i),\cdots,(g_n,G_n)\}$, let $D$ contain only the default graph.

\begin{enumerate}
    \item Let $V_{st}$ be the set of internal vertices in $V$. When iterating $V_{st}$, let $v_s$, $v_p$, $v_o$ be vertices with incoming $subject$, $predicate$, $object$ edges from a $v \in V_{st}$;
    \item For every $v \in V_{st}$ for which $\tau(v_p) \neq in$:
    \begin{enumerate}
        \item Create RDF-star triple $t=(s,p,o)$;
        \item Let $s=\tau(v_s)$, if $s \in S$ then execute steps (a), (b), (c), and (d) on $v_s$. Let $s$ be the RDF-star triple obtained in step (d);
        \item Let $p=\tau(v_p)$;
        \item Let $o=\tau(v_o)$, if $o \in S$ then execute steps (a), (b), (c), and (d) on $v_o$. Let $o$ be the RDF-star triple obtained in step (d);
        \item Let $V_{mem}$ be the set of vertices with outgoing edges to $v$;
        \item For all $v_{mem} \in V_{mem}$, let $\overline{v_o}$ be the vertex with incoming $object$ edge from $v_{mem}$:
        \begin{enumerate}
            \item Let $name = \tau(\overline{v_o})$;
            \item If $name = G_D$, add the triple $t$ to the default graph in $D$;
            \item If $name \neq G_D$, add the triple $t$ to the named graph named $name$ in $D$, if no such graph exists create a new one.
        \end{enumerate}
    \end{enumerate}
\end{enumerate}